# AUTOMATED MORPHOLOGICAL CLASSIFICATION OF APM GALAXIES BY SUPERVISED ARTIFICIAL NEURAL NETWORKS

A. Naim[1], O. Lahav[1], L. Sodré Jr.[2] & M. C. Storrie-Lombardi[1]

[1] *Institute of Astronomy, Madingley Rd., Cambridge, CB3 0HA, U.K.*
[2] *Instituto Astronômico e Geofísico da Universidade de São Paulo, CP9638, 01065-970, São Paulo, Brazil*

1 March 1995

**ABSTRACT**

We train *Artificial Neural Networks* to classify galaxies based solely on the morphology of the galaxy images as they appear on blue survey plates. The images are reduced and morphological features such as bulge size and the number of arms are extracted, all in a fully automated manner. The galaxy sample was first classified by 6 independent experts. We use several definitions for the mean type of each galaxy, based on those classifications. We then train and test the network on these features. We find that the rms error of the network classifications, as compared with the mean types of the expert classifications, is 1.8 Revised Hubble Types. This is comparable to the overall rms dispersion between the experts. This result is robust and almost completely independent of the network architecture used.

**Key words:**  galaxies: general - classification

## 1 INTRODUCTION

Since the introduction of the Hubble classification scheme (Hubble, 1926,1936) astronomers have been looking at ways to classify galaxies. Other systems were suggested, e.g. Mt. Wilson (Sandage 1961), Yerkes (Morgan 1958), Revised Hubble (de Vaucouleurs 1959), DDO (van den Bergh 1960a,b, 1976), and each has its special characteristics, but they all share Hubble's original notion that the sequence of morphologies attests to an underlying sequence of physical processes.

This notion has been widely accepted for the past few decades, making morphological classification of large numbers of galaxies important for better modelling and understanding of galaxy structure and evolution. Examples include statistical relations which are specific to certain types of galaxies, e.g. the $D_n - \sigma$ relation for ellipticals (Lynden Bell *et al.* 1988), the Tully-Fisher relation for spirals (Tully & Fisher 1977) and the morphology-density relation (Hubble 1936, Dressler 1980).

Morphological classification of galaxies is usually done by visual inspection of photographic plates. This is by no means an easy task, requiring skill and experience. It is also time consuming : Catalogues containing human classifications take years to complete and contain of order $10^4$ entries (e.g. The Third Reference Catalogue of Bright Galaxies (de Vaucouleurs *et al.* 1991); The ESO catalogue (Lauberts & Valentijn 1989)). However, in the APM (Automated Plate Measuring machine) survey (e.g. Maddox *et al.*, 1989) there are roughly $2 \times 10^6$ galaxies, and the expected yield of the Sloan Digital Sky Survey (Gunn *et al.*, in preparation) is over $10^7$ CCD images of galaxies. Clearly, such numbers of galaxies cannot be classified by humans. There is an obvious need for automated methods that will put the knowledge and experience of the human experts to use and produce very large samples of automatically classified galaxies.

The first stage towards achieving this goal was creating a uniform, well-defined sample to be classified by human experts. This was done in previous papers (Naim *et al.* 1994, hereafter paper I; Lahav *et al.* 1994), where the same sample of galaxies was presented to 6 independent expert observers and a detailed analysis of their classifications was carried out. The experts are : R. Buta, H. Corwin, G. de Vaucouleurs, A. Dressler, J. Huchra and S. van den Bergh (hereafter RB, HC, GV, AD, JH and vdB, respectively). We found that the rms dispersions between pairs of experts range from 1.3 to 2.1 Revised Hubble types, and that the overall rms dispersion was 1.8 types.

The next stage, which is carried out in this paper, entails training a computer software to classify galaxies on the basis of their apparent morphology. Our choice of an automated classifier is *Artificial Neural Networks* (ANNs), which proved in a pilot study (Storrie-Lombardi *et al.* 1992) to be well suited for this task. The original idea behind ANNs was creating a simplified model of the human brain (McCullogh & Pitts 1943, Hopfield & Tank 1986), but they were found to be well suited for a variety of applications in astronomy, such as classifying objects in the IRAS point source catalogue (e.g. Adorf & Meurs 1988), adaptive optics (e.g. Angel *et al.* 1990), scheduling observation time (e.g. Adorf 1989) and star-galaxy separation (e.g. Odewahn *et al.* 1991).

In this stage of our research we apply "Supervised Learning", whereby we attempt to teach the ANN to mimic the human classifications. The ANN is given a set of parameters describing each galaxy and is told what the "correct" type is. It then tries to make its classification as similar to the desired one as possible. (For more details on supervised learning, as well as other aspects of neural networks in the context of galaxy classification see Lahav *et al.* 1995).

The outline of this paper is as follows : In § 2 we discuss the galaxy sample. The various ways of defining mean galaxy types for training the ANN are explained in § 3. In § 4 we describe the preparation of galaxy images for feature extraction. The process of extracting morphological features for the ANN is described in detail in § 5. In § 6 we give the results of training various configurations of the ANN, based on different choices of input parameters, mean types and ANN architectures. The discussion follows in § 7. In the appendix we give a detailed listing of the human and ANN classifications, as well as of all the parameters the ANNs used, for a portion of the whole sample. The full table for the entire sample may be obtained from the authors.

## 2   DISCUSSION

We describe an automated process starting with the digitised images of galaxies and ending with their morphological classification. The sample was selected from the APM Equatorial Catalogue (§ 2), but the techniques described in this paper can be easily applied to other sources of galaxy images. Classifications for this sample were provided through collaborative efforts with six experts (§ 3), and can be used to automatically classify many other galaxies. The data reduction (§ 4) and feature extraction (§ 5) were done by our software. The ANN was trained on various sets of types and performed very well : The rms dispersion between the ANN types and the corrected mean type is comparable to the overall rms between pairs of experts (as found in paper I). Analysis of the way the ANN classifies (§ 6) showed that it is more or less reliable to the same degree for most types, but does not classify any galaxy into types $-5$, 9 or 10. Its error compared to the mean expert type is smallest for the type range $3-5$, which is the most frequent in the sample. These can be taken as measures of the ANN reliability in future classifications of many more images, predominantly of as yet unclassified galaxies. Another extension of this work can be the application of unsupervised methods to these data sets, which could result in a different, new classification scheme altogether.